\documentclass[a4paper]{article}
\usepackage[utf8]{inputenc}
\usepackage[english]{babel} 
\usepackage[noblocks]{authblk}
\usepackage{amsmath,amssymb,amsthm} 
\usepackage{graphicx}
\usepackage{esint}
\usepackage[color=yellow]{todonotes}
\usepackage{graphicx}
\usepackage{enumerate}
\usepackage{geometry}
\usepackage{float, tikz-cd, framed}
\usepackage{hyperref}
\usepackage{natbib}

\usepackage{ulem}

\usepackage{color}

\newcommand{\add}[1]{\textcolor{blue}{#1}}

\begin{document}

\title{An explicit method to determine Casimirs in 2D geophysical flows \thanks{This work was performed in the project SPRESTO (structure-preserving regularization and stochastic forcing for nonlinear hyperbolic PDEs), supported by a NWO TOP 1 grant.}}

\author[1]{Erwin Luesink}
\author[1,2]{Bernard Geurts}

\affil[1]{Multiscale Modelling and Simulation, Department of Applied Mathematics, Faculty EEMCS, University of Twente, PO Box 217, 7500 AE Enschede, The Netherlands}
\affil[2]{Multiscale Physics, Center for Computational Energy Research, Department of Applied Physics, Eindhoven University of Technology, PO Box 513, 5600 MB Eindhoven, The Netherlands}

\date{}
\maketitle

\begin{abstract}
Conserved quantities in geophysical flows play an important role in the characterisation of geophysical dynamics and aid the development of structure-preserving numerical methods. A significant family of conserved quantities is formed by the Casimirs i.e., integral conservation laws that are in the kernel of the underlying Poisson bracket. The Casimirs hence determine the geometric structure of the geophysical fluid equations among which the enstrophy is well known. Often Casimirs are proposed on heuristic grounds and later verified to be part of the kernel of the Poisson bracket. In this work, we will explicitly construct Casimirs by rewriting the Poisson bracket in vorticity-divergence coordinates thereby providing explicit construction of Casimirs for 2D geophysical flow dynamics.  
\end{abstract}

\section{Introduction}
Models of geophysical flows involve a fluid with a free interface under the influence of gravity on a rotating domain, often forced by variations in temperature, salinity and density. In this paper, we will consider the thermal rotating shallow water equations, which is a two dimensional model that includes all of the features above. The variations of temperature and salinity are collected in a single scalar field that we call the buoyancy. The thermal rotating shallow water equations can be derived in several different ways from the rotating Euler equations with stratification. The best known method is to assume that the domain is shallow, allowing the replacement of the prognostic equation for the vertical velocity by a hydrostatic pressure relation from which the vertical velocity can be inferred. The result is the system of inviscid primitive equations. Upon assuming columnar motion, e.g., motivated by the Taylor-Proudman theorem, meaning that the horizontal velocity field does not depend on the vertical coordinate, one can vertically average the primitive equations to obtain the thermal rotating shallow water equations. 

The focus in this contribution is with the preservation of key structures that characterise the governing equations and their dynamics. In geophysical fluid dynamics (GFD) models in 2D so-called conservation of Casimirs forms a major framework for the modelling and construction of methods for their numerical treatment. We present an explicit constructive method with which the Casimirs of a general class of GFD models can be explicitly computed. This provides an alternative to earlier work by \cite{cotter2013variational} that yields explicit Casimirs for 2D systems.

This entire sequence of assumptions and approximations can be performed at the level of the Lagrangian formulation of the models, as shown in \cite{holm2021stochastic}. Doing so is helpful because the geometric structure associated with such a variational formulation implies various important dynamical quantities and conservation laws. For two-dimensional geophysical fluids, we find that the key dynamical quantities are the potential vorticity and the potential buoyancy, where the latter is the ratio of the buoyancy to the depth. Both quantities play a key role in cyclogenesis, as shown in \cite{holm2021stochasticb}. In the present work, we will show that the potential vorticity and potential buoyancy can also be used to give an alternative formulation of the fluid equations. This formulation is particularly useful for the identification of Casimirs, which are conserved integral quantities. The Casimirs in turn help interpret mechanisms in geophysical turbulence. For instance, in the incompressible setting without buoyancy effects in two dimensions, the enstrophy is one of the Casimirs and plays a central role in the double cascade predicted by \cite{kraichnan1967inertial}. From a geometric point of view, the Casimirs are the functionals that form the kernel of the Poisson bracket. Usually, the form of the Casimirs is assumed or guessed and then verified by checking whether the Poisson bracket indeed vanishes. We will use two subsequent changes of variables to provide a constructive derivation for the form of the Casimirs.

The organization of this paper is as follows. In Section~\ref{geoflo} we present the main equations relevant to 2D geophysical flow and sketch the central importance of the retaining symmetries and conserved properties of the models. Among these properties, Casimirs are a major structure for the GFD models in 2D. Explicit methods to determine these Casimirs will be sketched in Section~\ref{casimirs}. Concluding remarks are collected in Section~\ref{conclu}.

\section{Geophysical flows}\label{geoflo}
In \cite{holm2021stochastic} it is shown that starting from the Lagrangian for the rotating, stratified Euler equations in three dimensions, one can derive the Lagrangian for the thermal rotating shallow water equations by assuming hydrostatic pressure and applying vertical averaging. The benefit of performing this sequence of approximations at the level of the Lagrangian is that the geometric structure is not affected. This means that the thermal rotating shallow water equations can be formulated using a Lie-Poisson bracket. As was shown in \cite{marsden1983coadjoint}, there exist multiple equivalent formulations in terms of Lie-Poisson brackets of the two-dimensional Euler equations. Similar formulations can be derived for the compressible two-dimensional fluid models. This was first done in \cite{holm1989lyapunov}. The retained Lagrangian structure in 2D is basic to a possible systematic approach to determining the conserved quantities associated with the thermal rotating shallow water equations. This systematic approach will be elaborated on in this paper.

It is possible to derive the equations for geophysical fluid dynamics on arbitrary smooth manifolds starting from the Euler equations, as shown in \cite{holm1998euler}. This level of generality requires tools of differential topology, such as the Lie derivative, the pullback and the pushforward, see for instance \cite{abraham1978foundations, marsden2013introduction, holm2009geometric}. However, if the domain is a two-dimensional compact subset $\Omega$ embedded in $\mathbb{R}^3$ or $\mathbb{R}^2$ with Cartesian coordinates and appropriate boundary conditions, we can formulate the equations of motion also using vector proxies of exterior calculus. The computations are in arbitrary orthonormal coordinate systems. To this end, we start from the dimensionless Lagrangian for the thermal rotating shallow water (TRSW) equations, which are given by
\begin{equation}\label{lag:trsw}
    \begin{aligned}
    L_{TRSW}(\boldsymbol{u},\eta,b) = \int_\Omega \left(\frac{1}{2}|\boldsymbol{u}|^2 + \frac{1}{{\rm Ro}}\boldsymbol{u}\cdot\boldsymbol{R} - \frac{1}{2\,{\rm Fr}^2}(1+\mathfrak{s}b)(\eta-2h)\right)\,\eta\,d\mu.
    \end{aligned}
\end{equation}
In this Lagrangian $\boldsymbol{u}$ is the velocity field, $\eta=\alpha\zeta + h$ is the total depth, $\zeta$ is the free surface elevation, $h$ is the bottom topography, $\boldsymbol{R}$ is the vector potential for the Coriolis parameter and $b$ is the buoyancy variable. Furthermore, the dimensionless numbers are the aspect ratio $\sigma$, the Rossby number ${\rm Ro}$, the Froude number ${\rm Fr}$, the wave amplitude $\alpha$ and the stratification parameter $\mathfrak{s}$. The stratification parameter represents the importance of the buoyancy variable, which itself is of order one. The vector potential for the Coriolis parameter satisfies $\nabla^\perp\cdot\boldsymbol{R} = f(x,y)$, where $f(x,y)$ is the usual Coriolis parameter. The $\perp$ operator corresponds to a Hodge star operator, which for Cartesian coordinates is defined by $(x,y)^\perp = (-y,x)$. The volume form is given by $d\mu$, which in Cartesian coordinates is expressed as $d\mu=dx\,dy$. The Lagrangian is a functional on $\mathfrak{X}(\Omega)\times V^*(\Omega)$, the product space of the space of vector fields $\mathfrak{X}(\Omega)$ and the space of advected quantities $V^*(\Omega)$. That is, $\boldsymbol{u}\in\mathfrak{X}(\Omega)$ and $\eta, b\in V^*(\Omega)$. More information and details on these spaces can be found in \cite{luesink2021stochastic}. The thermal rotating shallow water equations associated with the Lagrangian \eqref{lag:trsw} are given in the advective formulation by
\begin{equation}\label{eq:trsw_advective}
    \begin{aligned}
    \frac{\partial}{\partial t}\boldsymbol{u} + (\boldsymbol{u}\cdot\nabla)\boldsymbol{u} + \frac{1}{{\rm Ro}}f\boldsymbol{u}^\perp &= -\frac{\alpha}{{\rm Fr}^2}\nabla((1+\mathfrak{s}b)\zeta) + \frac{\mathfrak{s}}{2\,{\rm Fr}^2}(\alpha\zeta-h)\nabla b,\\
    \frac{\partial}{\partial t}\eta + \nabla\cdot(\eta\boldsymbol{u}) &= 0,\\
    \frac{\partial}{\partial t}b + (\boldsymbol{u}\cdot\nabla)b &= 0.
    \end{aligned}
\end{equation}
Equations \eqref{eq:trsw_advective} describe a compressible fluid with thermal effects in a rotating frame in two dimensions. Note that the equations take the same form on the sphere. The bold symbols denote vector valued quantities, which assumes the existence of a basis. Since we are working with two dimensional manifolds, such a basis is available. We will now recall an important vector calculus identity that is central to the manipulations that we will perform. This identity follows from differential topology, as shown in \cite{holm2009geometric}, where vector fields and differential forms in arbitrary coordinate systems can be defined. The vector-valued coefficients of these objects can be identified when the underlying space is Euclidean with the standard inner product. This means that $\boldsymbol{u}$ is simultaneously the coefficient of a vector field and a differential 1-form. This can lead to confusion, so in what follows $u$ denotes a vector field in arbitrary coordinates and $v$ denotes a differential 1-form. The Lie derivative can be defined as the derivative of the pullback
\begin{equation}\label{def:pullback}
    \mathcal{L}_u v = \frac{d}{dt}\Big|_{t=0}\phi_t^*(v),
\end{equation}
where $\phi$ is the flow associated with the vector field $u$. The asterisk on $\phi$ means that $v$ is pulled back along $\phi$. Here, $\phi$ is the solution to the differential equation $d\phi/dt = u(\phi,t)$ with arbitrary initial condition, i.e., representing a trajectory of a test-particle in the flow field $u$. The arbitrariness is what allows Lagrangian formulations of fluid dynamics to be related to Eulerian formulations. The Lie derivative can also be defined by means of Cartan's formula
\begin{equation}\label{def:cartan}
    \mathcal{L}_u v = i_u dv + d(i_u v),
\end{equation}
where $i_u v$ denotes the interior product of $u$ with $v$ and $d$ is the exterior derivative, see e.g. \cite{abraham1978foundations,marsden2013introduction,holm2009geometric}. Setting the two definitions \eqref{def:pullback} and \eqref{def:cartan} of the Lie derivative equal to one another in $\mathbb{R}^2$ provides the following vector calculus identity
\begin{equation}\label{eq:vectorcalculusidentity}
    (\boldsymbol{u}\cdot\nabla)\boldsymbol{v} + (\nabla\boldsymbol{u})\cdot\boldsymbol{v} = (\nabla^\perp\cdot\boldsymbol{v})\boldsymbol{u}^\perp + \nabla(\boldsymbol{u}\cdot\boldsymbol{v}),
\end{equation}
where the left hand side results from the pullback definition \eqref{def:pullback} and the right hand side follows from the Cartan formula \eqref{def:cartan}.

Using \eqref{eq:vectorcalculusidentity}, an alternative formulation to \eqref{eq:trsw_advective} of the thermal rotating shallow water equations is given by
\begin{equation}\label{eq:trsw_rot}
    \begin{aligned}
        \frac{\partial}{\partial t}\boldsymbol{u} + \left(\omega+\frac{1}{{\rm Ro}}f\right) \boldsymbol{u}^\perp &= -\frac{\alpha}{{\rm Fr}^2}\nabla((1+\mathfrak{s}b)\zeta) - \frac{1}{2}\nabla|\boldsymbol{u}|^2 + \frac{\mathfrak{s}}{2\,{\rm Fr}^2}(\alpha\zeta-h)\nabla b,\\
    \frac{\partial}{\partial t}\eta + \nabla\cdot(\eta\boldsymbol{u}) &= 0,\\
    \frac{\partial}{\partial t}b + (\boldsymbol{u}\cdot\nabla)b &= 0,\\
    \end{aligned}
\end{equation}
where $\omega=\nabla^\perp\cdot\boldsymbol{u}$ is the scalar vorticity. This formulation is often called the vector invariant formulation of fluid dynamics. There are several situations where the vector invariant formulation has advantages over the advective formulation. The vector invariant form offers an easier basis for numerical discretisations because it hides the nonlinearity. Additionally, on the right hand side of the velocity equation in \eqref{eq:trsw_rot}, one recognises the gradient of the Bernoulli function $B=\frac{1}{2}|\boldsymbol{u}|^2 + \frac{\alpha}{{\rm Fr}^2}(1+\mathfrak{s}b)\zeta$, further explaining the physical mechanisms responsible for the flow of fluid. 

Both \eqref{eq:trsw_advective} and \eqref{eq:trsw_rot} can be formulated as a Hamiltonian system with a Lie-Poisson bracket. This is a crucial property of these models as it prepares for a clear formulation of conserved quantities, i.e., the desired Casimirs. By Legendre transforming the Lagrangian \eqref{lag:trsw}, we obtain the Hamiltonian associated with the thermal rotating shallow water equations. The Legendre transform is given by
\begin{equation}
    H(\boldsymbol{m},\eta,b) = \langle \boldsymbol{m},\boldsymbol{u}\rangle - L(\boldsymbol{u},\eta,b),
\end{equation}
where $\boldsymbol{m}=\delta L/\delta\boldsymbol{u}$ is the momentum variable in terms of the functional derivative of the Lagrangian. The Legendre transform only relates the velocity and the momentum and does not affect the advected quantities. The Hamiltonian is therefore a functional on the space $\mathfrak{X}^*(\Omega)\times V^*(\Omega)$, where $\mathfrak{X}^*(\Omega)$ is the dual space of $\mathfrak{X}(\Omega)$ with respect to $L^2$-pairing on $\Omega$. For the thermal rotating shallow water model, the momentum is given by $\boldsymbol{m}=\eta(\boldsymbol{u}+\boldsymbol{R})$ and the Hamiltonian is
\begin{equation}
    H_{TRSW}(\boldsymbol{m},\eta,b) = \int_\Omega \left(\frac{1}{2\eta^2}|\boldsymbol{m}|^2 + \frac{1}{2\,{\rm Fr}^2}(1+\mathfrak{s}b)(\eta-2h)\right)\,\eta\,d\mu.
\end{equation}
The Poisson bracket that will allow us to formulate the equations of motion has the following general form, see e.g. \cite{holm2009geometric},
\begin{equation}
    \{F,G\} = \int_\Omega \left(\begin{matrix} 
    \delta F/\delta \boldsymbol{m} \\
    \delta F/\delta \eta\\
    \delta F/\delta b
    \end{matrix}\right)\cdot \mathbb{J}(\boldsymbol{m},\eta,b) \left(\begin{matrix} 
    \delta G/\delta \boldsymbol{m} \\
    \delta G/\delta \eta\\
    \delta G/\delta b
    \end{matrix}\right) \,d\mu\,,
\end{equation}
where $F$ and $G$ are two arbitrary functionals on the space $\mathfrak{X}^*(\Omega)\times V^*$. The matrix $\mathbb{J}$ is operator valued and depends on the state. $\mathbb{J}$ can be explicitly derived using reduction theorems, following \cite{marsden1974reduction, holm1998euler}. We speak of Hamiltonian dynamics if the evolution of a functional $F$ is expressed by
\begin{equation}
    \frac{d}{dt}F = -\{F,H\},
\end{equation}
where $H$ is the Hamiltonian of the system. We obtain the equations of motion \eqref{eq:trsw_advective} by choosing $G=H$ and using $\mathbb{J}(\boldsymbol{m},\eta,b)$ as in \cite{holm2021stochasticb} to find
\begin{equation}\label{eq:lpbracket_advective} 
    \frac{\partial}{\partial t}\left(\begin{matrix}
    m_i\\
    \eta\\
    b
    \end{matrix}\right) = \underbrace{\left(\begin{matrix}
    m_j\partial_i + \partial_j m_i & \eta\partial_i & -b_{,i}\\
    \partial_j\eta & 0 & 0\\
    b_{,j} & 0 & 0
    \end{matrix}\right)}_{=\mathbb{J}(\boldsymbol{m},\eta,b)}\left(\begin{matrix}
    \delta H_{TRSW}/\delta m_j\\
    \delta H_{TRSW}/\delta \eta\\
    \delta H_{TRSW}/\delta b
    \end{matrix}\right)
\end{equation}
and reproduces the equations \eqref{eq:trsw_advective} after using the definition of the momentum, rewriting the momentum equation using the continuity equation and applying the Lie derivative identity \eqref{eq:vectorcalculusidentity} to the vector potential of the Coriolis parameter. We have used Einstein's summation convention of summing over repeated indices and the notation $b_{,i}$ refers to $i$-th component of the gradient of $b$ (the comma denotes a spatial derivative). Note that the dependence on the state $(\boldsymbol{m},\eta,b)$ is linear. This means that the Poisson bracket is in fact a Lie-Poisson bracket. The kernel of a Lie-Poisson bracket is nontrivial and is precisely the kernel of the linear operator $\mathbb{J}$. The kernel is spanned by functionals known as Casimirs and the goal of the remainder of this work is to derive these Casimir functionals explicitly.

In \cite{holm1989lyapunov}, the momentum bracket \eqref{eq:lpbracket_advective} is transformed by means of a change of variables to the rotational form. The change of variables is $(m_1,m_2,\eta,b)\mapsto (u,v,\eta,b)$, where $u=\frac{m_1}{\eta}$ and $v=\frac{m_2}{\eta}$, which is invertible as long as $0< \eta < \infty$. This change of variables is not the same as just formulating the Hamiltonian and the Lie-Poisson bracket in terms of velocity because of the role of the rotating frame. Performing the transformation of \cite{holm1989lyapunov}, we obtain the following formulation
\begin{equation}\label{eq:lpbracket_rot}
    \frac{\partial}{\partial t}\left(\begin{matrix}
    u\\
    v\\
    \eta\\
    b
    \end{matrix}\right) = \underbrace{\left(\begin{matrix}
    0 & -q & \partial_x & -\frac{b_x}{\eta}\\
    q & 0 & \partial_y & -\frac{b_y}{\eta}\\
    \partial_x & \partial_y & 0 & 0\\
    \frac{b_x}{\eta} & \frac{b_y}{\eta} & 0 & 0
    \end{matrix}\right)}_{=\mathbb{J}(u,v,\eta,b)}\left(\begin{matrix}
    \delta H_{TRSW}/\delta u = \eta u\\
    \delta H_{TRSW}/\delta v = \eta v\\
    \delta H_{TRSW}/\delta \eta = B\\
    \delta H_{TRSW}/\delta b = T
    \end{matrix}\right),
\end{equation}
where $q = \frac{1}{\eta}(\omega + \frac{1}{{\rm Ro}}f)$ is the potential vorticity, $B = \frac{1}{2}|\boldsymbol{u}|^2 + \frac{1}{2\,{\rm Fr}^2}(1+\mathfrak{s}b)(\eta-2h)$ is the Bernoulli function and $T=\frac{\mathfrak{s}}{2\,{\rm Fr}^2}(\eta^2 - 2\eta h)$. This Hamiltonian formulation corresponds to \eqref{eq:trsw_rot} and extends the bracket of \cite{holm1989lyapunov} to include thermal variations. This bracket is linear and it is a simple exercise to show that it is skew-symmetric with respect to the $L^2$-pairing. To prove that it satisfies the Jacobi identity, we repeat the argument of \cite{holm1989lyapunov} and state that the bracket in \eqref{eq:lpbracket_rot} is an invertible transformation of variables in the Lie-Poisson bracket \eqref{eq:lpbracket_advective}.

We now wish to express the bracket \eqref{eq:lpbracket_rot} in another set of variables to achieve a formulation in independent scalar variables. This is accomplished using the transformation $(u,v,\eta,b)\mapsto (\omega,D,\eta,\eta b)$, where $\omega = \nabla^\perp\cdot \boldsymbol{u}$ is the vorticity and $D=\nabla\cdot\boldsymbol{u}$ is the divergence. If the domain $\Omega$ has no holes, then the vector Laplace equation has a unique solution. This is the Helmholtz theorem for Euclidean space. If the domain does have holes (or islands) then appropriate boundary conditions are required to eliminate harmonic functions in order to guarantee uniqueness of solutions. In either case, provided with appropriate boundary conditions when necessary, one can uniquely split up the vector field $\boldsymbol{u}$ into potentials via the Helmholtz decomposition
\begin{equation}
    \boldsymbol{u} = \nabla^\perp\psi + \nabla\chi,
\end{equation}
where $\psi$ is the stream function and $\chi$ is the velocity potential. Uniqueness of solutions to the Laplace equation is required to reconstruct $\boldsymbol{u}$, since the potentials satisfy Poisson's equations 
\begin{equation}
    \begin{aligned}
    \omega &= \nabla^\perp\cdot(\nabla^\perp\psi) &= \Delta\psi,\\
    D &= \nabla\cdot(\nabla\chi) &= \Delta\chi.
    \end{aligned}
\end{equation}
This permits us to write the velocity field in terms of vorticity and divergence as follows
\begin{equation}\label{eq:helmholtz}
    \boldsymbol{u} = \nabla^\perp\Delta^{-1}\omega + \nabla\Delta^{-1}D,
\end{equation}
which means that changing variables in the Hamiltonian amounts to substituting \eqref{eq:helmholtz} for the velocity. The first term in \eqref{eq:helmholtz} is analogous to the two-dimensional version of the Biot-Savart law, but since the fluid is compressible it does not determine the full velocity field. The potential buoyancy variable $r = b/\eta$ is introduced to remove the fraction terms $\pm b_x/\eta$ and $\pm b_y/\eta$ in the Poisson bracket \eqref{eq:lpbracket_rot}. This is convenient from a notation point of view, but also means that all the variables that appear inside the $\mathbb{J}$ matrix in the Poisson bracket are ``potential variables", i.e., potential vorticity and potential buoyancy. To change variables in the Cartesian coordinate case, we sandwich the Poisson bracket of equation \eqref{eq:lpbracket_rot} with the functional Jacobian derivative operator and its $L^2$-adjoint as follows
\begin{equation}
    \left(\begin{matrix}
    -\partial_y & \partial_x & 0 & 0\\
    \partial_x & \partial_y & 0 & 0\\
    0 & 0 & 1 & 0\\
    0 & 0 & 0 & 1
    \end{matrix}\right)
    \left(\begin{matrix}
    0 & -q & \partial_x & -\frac{b_x}{\eta}\\
    q & 0 & \partial_y & -\frac{b_y}{\eta}\\
    \partial_x & \partial_y & 0 & 0\\
    \frac{b_x}{\eta} & \frac{b_y}{\eta} & 0 & 0
    \end{matrix}\right)
    \left(\begin{matrix}
    \partial_y & -\partial_x & 0 & 0\\
    -\partial_x & -\partial_y & 0 & 0\\
    0 & 0 & 1 & 0\\
    0 & 0 & 0 & 1
    \end{matrix}\right)
\end{equation}
Similar manipulations can be performed for different coordinate systems, provided that one takes an orthonormal coordinate basis. For the sphere this would be a latitude-longitude basis. Performing this change of coordinates in the bracket \eqref{eq:lpbracket_rot} leads to 
\begin{equation}\label{eq:lpbracket_magic}
\Big\{ F, G\Big\}
=
- \int_\Omega\left(
\begin{matrix}
\delta F/ \delta \omega \\  
\delta F/ \delta D \\ 
\delta F/ \delta \eta \\ 
\delta F/ \delta b
\end{matrix}\right)
\nabla
\cdot
\underbrace{
\left(\begin{matrix}
q\,\times & -q & 0 & r\,\times \\
q & q\,\times & 1 & r \\
0 & -1 & 0 & 0 \\
r\,\times & -r & 0 & 0
\end{matrix}\right)}_{=\mathbb{J}(\omega,D,\eta,b)}
\nabla
\left(
\begin{matrix}
\delta G / \delta \omega \\
\delta G / \delta D \\
\delta G / \delta \eta \\
\delta G / \delta b
\end{matrix}\right)
\,d\mu\,.
\end{equation}
This bracket formulation has a number of interesting properties. First of all, the differential operators can be factored out of the matrix upon introducing the convention $\times\nabla = \nabla^\perp$. Secondly, the matrix only features the potential vorticity $q$ and the potential buoyancy $r$. The kernel of this bracket is the kernel of the matrix operator $\mathbb{J}(\omega,D,\eta, b)$, which is skew-symmetric in both $q$ and $r$. The most interesting property in our opinion is that this bracket governs a number of two-dimensional fluid models through submatrices of $\mathbb{J}$. 

The full four by four setting corresponds to the thermal rotating shallow water equations. If there is no underlying rotation, one simply adapts the definition of the potential vorticity $q$ to obtain the thermal shallow water equations. When there are no buoyancy variations, the bracket can be restricted to the three by three case, which corresponds to the rotating shallow water equations. Again, upon adapting the potential vorticity variable $q$, one can obtain the shallow water equations in case rotation is absent. In the incompressible case where buoyancy variations still play a role, the divergence is zero. Then one can use the submatrix consisting of the $(1,1),(1,4),(4,1),(4,4)$ elements, which corresponds to the thermal rotating Euler equations. If buoyancy variations do not play a role, the matrix is simply the $(1,1)$ element, which can describe the two\add{-}dimensional rotating Euler equations and the quasi-geostrophic (QG) equations. Important to note is that the thermal QG (TQG) model described in \cite{warneford2013quasi, zeitlin2018geophysical, holm2021stochasticb} does not fit into the Lie-Poisson bracket formulation because its $(1,1)$ position features $q-b$ rather than just $q$, see \cite{holm2021stochasticb}. This is a result of the fact that the TQG model is derived as a perturbation around thermal geostrophic balance, rather than around geostrophic balance. To summarise, we list all models that can be described by \eqref{eq:lpbracket_magic} in order of complexity in Table \ref{tab:pb}.

\begin{table}[h!]
\centering
\begin{tabular}{ll|ll}
    & Incompressible & & Compressible\\
    \hline
    1. & Euler                      & 5. & Shallow water\\
    2. & Rotating Euler             & 6. & Rotating shallow water\\
    3. & QG                         & 7. & Thermal shallow water\\
    4. & Thermal rotating Euler     & 8. & Thermal rotating shallow water
\end{tabular}
\caption{The models that can be described by submatrices of the Poisson bracket \eqref{eq:lpbracket_magic}.}
\label{tab:pb}
\end{table}
The transition from incompressible models to compressible models is a steep increase in complexity, since the compressible models involve an additional two equations compared to the incompressible case. The divergence and the depth variable are always paired together, since changes in $D$ imply changes in $\eta$ and vice-versa. This bracket is particularly useful in the explicit computation of Casimirs, which is shown in section \ref{casimirs}. It also has other uses. In ocean dynamics, gravity waves propagate at speeds that are orders of magnitude higher than the typical flow velocity. If the Rossby number, the Froude number, the wave amplitude and the stratification parameter satisfy $\mathcal{O}({\rm Ro}) = \mathcal{O}({\rm Fr}) = \mathcal{O}(\alpha) = \mathcal{O}(\mathfrak{s})$, then one can derive the thermal geostrophic balance condition. This condition provides an algebraic relation for the balanced velocity field. The balanced velocity field is divergence free. Since the bracket \eqref{eq:lpbracket_magic} features the divergence variable explicitly, it is natural to perform an asymptotic expansion in a small parameter $\epsilon$ where
\begin{equation}\label{eq:asymptoticexpansion}
\begin{aligned}
    \omega &= \omega_0 + \epsilon\omega_1 + o(\epsilon),\\
    D &= \epsilon D_1 + o(\epsilon).
\end{aligned}
\end{equation}
The expansion \eqref{eq:asymptoticexpansion} applied to the thermal rotating shallow water Lagrangian \eqref{lag:trsw} and truncated at order $o(1)$ yields the thermal extension of the L1 model of \cite{salmon1983practical}. This derivation is shown in detail in \cite{holm2021stochasticb}.

\section{Explicitly determining the Casimirs}\label{casimirs}
The bracket \eqref{eq:lpbracket_magic} is particularly helpful in the explicit computation of the Casimirs since the variables are all independent scalars. We are looking for functionals $C(\omega,D,\eta, b)$ such that $\{F,C\}=0$ for any functional $F$. The following computations follow a procedure of step-by-step elimination. Expanding the bracket and requiring $\{F,C\}=0$ for any $F$ yields the equation
\begin{equation}\label{eq:casimirsstep1}
    \begin{aligned}
    0 = \{F,C\} &= -\int_\Omega \frac{\delta F}{\delta\omega}\nabla\cdot\left(q\nabla^\perp\frac{\delta C}{\delta \omega} - q\nabla\frac{\delta C}{\delta D} +r\nabla^\perp\frac{\delta C}{\delta b}\right) 
    \\&\quad + \frac{\delta F}{\delta D}\nabla\cdot\left(q\nabla\frac{\delta C}{\delta \omega} + q\nabla^\perp\frac{\delta C}{\delta D} + \nabla\frac{\delta C}{\delta \eta} + r\nabla\frac{\delta C}{\delta b}\right)
    \\&\quad -\frac{\delta F}{\delta \eta}\nabla\cdot\left(\nabla\frac{\delta C}{\delta D}\right)
    \\&\quad +\frac{\delta F}{\delta b}\nabla\cdot\left(r\nabla^\perp\frac{\delta C}{\delta\omega} - r\nabla\frac{\delta C}{\delta D}\right) \, d\mu\,.
    \end{aligned}
\end{equation}
Since $F$ is an arbitrary functional, we can solve for this functional equation per variational derivative of $F$ and get the explicit form of the Casimirs by a process of elimination. In fact, the third line of \eqref{eq:casimirsstep1}, the one that features $\delta F/\delta \eta$, will be our starting point in the explicit computations. The third line implies that $C$ may be at most linear in $D$ with constant coefficients, since only then the variational derivative of $C$ with respect to $D$ is constant in space. Applying the gradient leads to zero, meaning that the third term vanishes under the assumption of linear dependence of $C$ on mass density $D$. If the variational derivative of $C$ with respect to $D$ is not constant with respect to space, then this term does not vanish, as $\Delta D$ is not necessarily zero. So at this stage we know that 
\begin{equation}
    C(\omega,D,\eta,b) = \int_\Omega \gamma D + f(\omega,\eta,b)\,d\mu\,,
\end{equation}
where $\gamma\in\mathbb{R}$ is a constant and $f(\omega,\eta,b)$ is a function that is determined next. Since we have established that $C$ must be linear in $D$, all terms that involve variational derivatives of $C$ with respect to $D$ vanish. Simplifying \eqref{eq:casimirsstep1} accordingly, we obtain
\begin{equation}\label{eq:casimirsstep2}
    \begin{aligned}
    0 = \{F,C\} &= -\int_\Omega \frac{\delta F}{\delta\omega}\nabla\cdot\left(q\nabla^\perp\frac{\delta C}{\delta \omega} +r\nabla^\perp\frac{\delta C}{\delta b}\right) 
    \\&\quad + \frac{\delta F}{\delta D}\nabla\cdot\left(q\nabla\frac{\delta C}{\delta \omega} + \nabla\frac{\delta C}{\delta \eta} + r\nabla\frac{\delta C}{\delta b}\right)
    \\&\quad +\frac{\delta F}{\delta b}\nabla\cdot\left(r\nabla^\perp\frac{\delta C}{\delta\omega}\right) \, d\mu\,.
    \end{aligned}
\end{equation}
The third line of \eqref{eq:casimirsstep2}, the one that features $\delta F/\delta b$, vanishes if $C$ depends linearly on $\omega$. In this case we do not have to insist on constant coefficients. If the coefficient of $\omega$ is an arbitrary differentiable function $\Psi(r)$ of the potential buoyancy, i.e., $\delta C/\delta \omega =\Psi(r)$, we have
\begin{equation}
    \nabla\cdot\left(r\nabla^\perp\frac{\delta C}{\delta\omega}\right) = \nabla\cdot\big(r\nabla^\perp \Psi(r)\big) =  \Psi'(r)(\nabla r\cdot\nabla^\perp r) + r(\nabla\cdot\nabla^\perp) \Psi(r) = 0,
\end{equation}
because $\nabla r, \nabla^\perp r$ and $\nabla,\nabla^\perp$ are orthogonal. Here $\Psi'(r)$ denotes the derivative of $\Psi$ with respect to its argument. Recall that the potential buoyancy is defined as $r=b/\eta$. At this stage, we know that $C$ must have the form
\begin{equation}
    C(\omega,D,\eta,b) = \int_\Omega \gamma D + \omega \Psi(r) + g(b,\eta)\,d\mu\,.
\end{equation}
The next step is to determine $g(b,\eta)$. Knowing the explicit form of the variational derivative of $C$ with respect to $\omega$, we can simplify \eqref{eq:casimirsstep2} further to obtain
\begin{equation}\label{eq:casimirsstep3}
    \begin{aligned}
    0 = \{F,C\} &= -\int_\Omega \frac{\delta F}{\delta\omega}\nabla\cdot\left(q\nabla^\perp\Psi(r) +r\nabla^\perp\frac{\delta C}{\delta b}\right) 
    \\&\quad + \frac{\delta F}{\delta D}\nabla\cdot\left(q\nabla\Psi(r) + \nabla\frac{\delta C}{\delta \eta} + r\nabla\frac{\delta C}{\delta b}\right)\, d\mu\,.
    \end{aligned}
\end{equation}
We focus on the term that features $\delta F/\delta \omega$. We have an explicitly constructed term and the variational derivative of $C$ with respect to $b$. By the same argument as in the previous step, we know $\nabla\cdot(r\nabla^\perp(\delta C/\delta b))$ vanishes if the variational derivative of $C$ with respect to $b$ is a differentiable function of $r$. So, let $\Phi(r)=\Phi(b/\eta)$ be this differentiable function. Since a variational derivative of $\Phi(b/\eta)$ with respect to $b$ produces a factor of $1/\eta$, we introduce the term $\eta\Phi(b/\eta)$ into $C$. So at this stage, after the three steps, we have an expression for $C$ in terms of two arbitrary differentiable functions $\Phi$ and $\Psi$
\begin{equation}\label{eq:casimirs}
C(\omega,D,\eta,b) = \int_\Omega \eta\Phi(r) + \eta q \Psi(r) + \gamma D\,d\mu\,  
\end{equation}
where $\gamma\in\mathbb{R}$ is a constant and $\eta q = \omega + \frac{1}{{\rm Ro}}f$. Going back to \eqref{eq:casimirsstep3}, we can verify whether $C(\omega,D,\eta,b)$ is indeed the family of Casimirs of the Poisson bracket \eqref{eq:lpbracket_magic}. This means that the terms multiplying variational derivatives of $F$ must vanish. So we substitute \eqref{eq:casimirs} into \eqref{eq:casimirsstep3} and compute the term multiplied by $\delta F/\delta \omega$
\begin{equation}
    \begin{aligned}
    \nabla\cdot\left(q\nabla^\perp\frac{\delta C}{\delta\omega}\right. &- \left.q\nabla\frac{\delta C}{\delta D} + r\nabla^\perp\frac{\delta C}{\delta b}\right) 
    = \nabla\cdot\left(q\nabla^\perp\Psi(r) + r\nabla^\perp\big(\Phi'(r) + q\Psi'(r)\big)\right) \\
    &= \Phi''(r)(\nabla r\cdot \nabla^\perp r) + r(\nabla\cdot\nabla^\perp) \Phi'(r) + \Psi'(r)(\nabla q\cdot \nabla^\perp r) + \Psi'(r)(\nabla r\cdot\nabla^\perp q)
    \\&\quad + r\Psi''(r)(\nabla r\cdot\nabla^\perp q) + r\Psi''(r)(\nabla q\cdot\nabla^\perp r)\\
    &= 0.
    \end{aligned}
\end{equation}
In this computation we have used orthogonality of $\nabla, \nabla^\perp$ and $\nabla r,\nabla^\perp r$ and skew-symmetry, i.e., $\nabla r\cdot\nabla^\perp q = -\nabla q\cdot\nabla^\perp r$ ($ = -r_xq_y + q_xr_y$ for Cartesian coordinates). It remains to check whether the term multiplied by $\delta F/\delta D$ also vanishes. In the following computation, we suppress the dependence of $\Phi$ and $\Psi$ on $r$ for notational convenience. We compute from the second line in \eqref{eq:casimirsstep1}
\begin{equation}\label{eq:casimirsstep4}
    \begin{aligned}
    \nabla\cdot\left(q\nabla\frac{\delta C}{\delta\omega} \right. &+ \left. q\nabla^\perp\frac{\delta C}{\delta D}+ \nabla\frac{\delta C}{\delta \eta} + r\nabla\frac{\delta C}{\delta b}\right) = \nabla\cdot\Big(q\nabla\Psi + \nabla\big(\Phi-r\Phi'-rq\Psi'\big) +  r\nabla(\Phi'+q\Psi') \Big)
    \\&\quad = \nabla\cdot\Big(q\nabla\Psi - r\nabla(q\Psi') - q\Psi'\nabla r + r\nabla(q\Psi')\Big)
    \\&\quad = 0.
    \end{aligned}
\end{equation}
The computation is a sequence of applying the identity $\nabla f(r) = f'(r)\nabla r$. In the first step we have applied this to all the terms that involve $\Phi$. Performing the same manipulations on $\Psi$ subsequently, implies the result. Note that in \eqref{eq:casimirsstep3} we required cancellations of terms orthogonal to the terms that are required to cancel in \eqref{eq:casimirsstep4}. The reason that the perpendicular gradient of the variational derivative of $C$ with respect to $\eta$ does not appear in \eqref{eq:casimirsstep3} is because it is trivially zero due to orthogonality of $\nabla$ and $\nabla^\perp$. Hence we can conclude that \eqref{eq:casimirs} is the complete description of the Casimirs for the bracket \eqref{eq:lpbracket_magic} and thus also for \eqref{eq:lpbracket_advective}. We can repeat the computation for each of the models described in Table \ref{tab:pb} to obtain the corresponding Casimirs. This is summarised in Table \ref{tab:casimirs}. Here one can see that in presence of thermal effects enstrophy is no longer a Casimir.

\begin{table}[h!]
\centering
\begin{tabular}{lll}
    & Incompressible & Casimirs \\
    \hline
    1. & Euler                      & $C= \int_\Omega \Phi(q) \,d\mu$ \\ 
    2. & Rotating Euler             & $C= \int_\Omega \Phi(q) \,d\mu$ \\ 
    3. & QG                         & $C= \int_\Omega \Phi(q) \,d\mu$ \\ 
    4. & Thermal rotating Euler     & $C= \int_\Omega \Phi(r) + q\Psi(r)\,d\mu$ \\
    & & \\
    & Compressible & Casimirs \\
    \hline
    5. & Shallow water                    & $C= \int_\Omega \eta\Phi(q) + \gamma D\,d\mu$\\
    6. & Rotating shallow water           & $C= \int_\Omega \eta\Phi(q) +\gamma D\,d\mu$\\
    7. & Thermal shallow water            & $C= \int_\Omega \eta\Phi(r) + \eta q\Psi(r) + \gamma D\,d\mu$\\
    8. & Thermal rotating shallow water   & $C= \int_\Omega \eta\Phi(r) + \eta q\Psi(r) + \gamma D\,d\mu$
\end{tabular}
\caption{The Casimirs of the models that can be described by submatrices of the Poisson bracket \eqref{eq:lpbracket_magic}.}
\label{tab:casimirs}
\end{table}

\section{Conclusion}\label{conclu}
In geometric approaches to fluid dynamics one often may exploit constructive methods to infer conservation laws via Noether's theorem \cite{abraham1978foundations, marsden1983coadjoint, marsden2013introduction, holm2009geometric}. Casimirs identify conservation laws that arise as the kernel of the Poisson bracket. We provided an explicit method of determining these Casimir functionals for two-dimensional fluid dynamics by means of two changes of variables for the thermal rotating shallow water equations. We formulated the equations in vector invariant form by using an important vector calculus identity. The Poisson bracket corresponding to this vector invariant form is convenient for further manipulation. By changing coordinates from velocity to vorticity and divergence, we derive a Poisson bracket that only involves (skew) gradients and divergences. By means of this formulation, it is a systematic computation to obtain the Casimirs. The computations were performed for arbitrary coordinate systems, which means that the above computations can easily used for domains such as the sphere.

\section*{Acknowledgements}
The authors are grateful for many fruitful discussions with Paolo Cifani, Sagy Ephrati, Arnout Franken and Darryl Holm. We thank the anonymous reviewer for their valuable comments.

\bibliographystyle{plainnat}
\bibliography{biblio.bib}

\end{document}